\documentclass[twocolumn,preprintnumbers]{revtex4}
\usepackage{amsmath}
\usepackage{graphicx}
\usepackage{dcolumn}
\usepackage{bm}

\begin{document}

\title{
	Transverse ``resistance overshoot" in a Si/SiGe two-dimensional electron gas\\
	in the quantum Hall effect regime}
%\shorttitle{Transverse ``resistance overshoot"\dots}	
\author{I.~Shlimak$^1$, V.~Ginodman$^1$, A.~B.~Gerber$^2$,
A.~Milner$^2$, K.-J.~Friedland$^3$, and D.~J.~Paul$^4$}
%\shortauthor{I. Shlimak \etal}
%\institute{
\affiliation{
  $^1$Minerva Center and Jack and Pearl Resnick Institute of Advanced Technology,
 	Department of Physics, Bar-Ilan University, 52900 Ramat-Gan, Israel\\
  $^2$School of Physics {\em\&}
  	Astronomy, Raymond {\em\&} Beverly Sackler Faculty of Exact Sciences,
	Tel-Aviv University, 69978 Tel Aviv, Israel\\
  $^3$Paul-Drude-Institut f\"{u}r Festk\"{o}rperelektronik,
	Hausvogteiplatz 5-7, 10117, Berlin, Germany\\
  $^4$Cavendish Laboratory, University of Cambridge, Madingley Road,
	Cambridge CB3 0HE, U.K.
}
%\pacs{73.43.-f}{Quantum Hall effect}
%\pacs{72.20.Ee}{Mobility edges, hopping transport}

\begin{abstract}
We investigate the peculiarities of the ``overshoot'' phenomena in the
transverse Hall resistance $R_{xy}$ in Si/SiGe. Near the
low magnetic field end of the quantum Hall effect plateaus, when the filling
factor $\nu$ approaches an integer $i$, $R_{xy}$ overshoots the normal
plateau value $h/ie^{2}$. However, if magnetic field $B$ increases further,
$R_{xy}$ decreases to its normal value. It is shown that in the investigated
sample $n$-Si/Si$_{0.7}$Ge$_{0.3}$, overshoots exist for almost all $\nu$.
Existence of overshoot in $R_{xy}$ observed in different materials and for
different $\nu$, where splitting of the adjacent Landau bands has different
character, hints at the common origin of this effect. Comparison of the
experimental curves $R_{xy}(\nu )$ for $\nu =3$ and $\nu =5$ with and
without overshoot showed that this effect exist in the whole interval
between plateaus, not only in the region where $R_{xy}$ exceeds the normal
plateau value.
\end{abstract}

\maketitle

Observations have been reported~\cite{b01,b02,b03,b04,b05} of anomalous
peaks in the Hall resistance $R_{xy}$ in high mobility GaAs/AlGaAs
heterojunctions measured in the quantum Hall effect regime. In incremental
magnetic fields $B$, when the filling factor $\nu$ approaches an odd
integer $i$, $R_{xy}$ overshoots the normal plateau value $h/ie^{2}$.
However, if $B$ increases further, $R_{xy}$ decreases to its normal value.
It was mentioned that in GaAs/AlGaAs heterostructures, the overshoots occur
near the low magnetic field end of the spin resolved odd plateaus. The
explanatory model was based on the assumption that these anomalies are due
to the decoupling of the two edge states of the topmost spin-split Landau
bands (LB), which occurs with an increase of the magnetic field because of
the enhancement of the $g$-factor and the corresponding spin splitting due
to the electron-electron interaction~\cite{b05}. Subsequently, this effect
was also observed in $n$- and $p$-Si/SiGe heterostructures~\cite{b06,b07,b08,b09}
where the above explanation is questionable. Indeed, for
Si-based heterostructures, the $g$-factor is initially large. Moreover, in
$n$-type structures, the overshoot was observed near $i = 3$, where spin
splitting is not relevant, because at odd $i$, adjacent LBs in $n$-type
Si/SiGe are valley-split. This hints at a more universal character of the
overshoot.

In this work we present the results of experimental investigations of the
overshoot in $n$-type Si/Si$_{1-x}$Ge$_x$ heterostructure. The sample
investigated was Hall-bar patterned $n$-type Si/Si$_{0.7}$Ge$_{0.3}$ double
heterostructure, 7~nm $i$-Si quantum well was situated between 1~$\mu$m
$i$-Si$_{0.7}$Ge$_{0.3}$ layer and 67~nm Si$_{0.7}$Ge$_{0.3}$ layer with
17~nm spacer followed by 50~nm Si$_{0.7}$Ge$_{0.3}$ heavily doped with
As. A 4~nm silicon cap layer protects the surface. The electron
concentration $n$ and mobility $\mu$ at 1.5~K were $n_0 = 9\cdot
10^{15}$~m$^{-2}$, $\mu_0 = 8$~$m^2$/V$\cdot$ s. The sample resistance was
measured using a standard lock-in-technique, with the measuring current
being 20~nA at a frequency of 10.6~Hz. The results of measurements
weakly depend on the choice of contacts of similar geometry. The results of
investigation of longitudinal conductivity in this sample were published
in~\cite{b10,b11}.

\begin{figure}[b]
\includegraphics[width=7 cm]{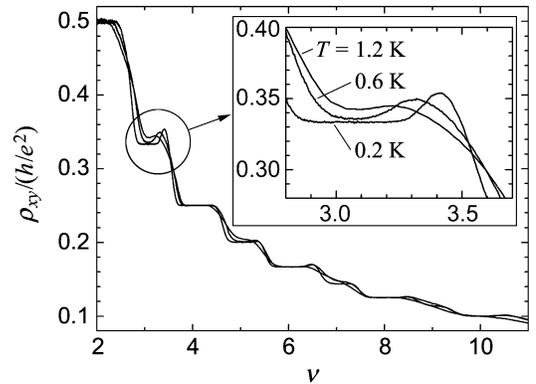}
\caption{Transverse Hall resistivity $\rho _{xy}$ in units of
$h/e^{2}$ as a function of filling factor $\nu$ measured at $T=0.2$,
0.6 and 1.2~K. The insert shows the enhanced view of the overshoot for
$\nu =3$.}
\label{f.1}
\end{figure}

Figure~\ref{f.1} shows the transversal Hall resistivity $\rho _{xy}$
measured in units of $h/e^2=25.8$~k$\Omega$ at different temperatures 0.2,
0.6 and 1.2~K as a function of filling factor $\nu =n_{0}h/eB$. The
enhanced view of the overshoots near the vicinity of $\nu =3$ is shown in
the inserts. The maximal amplitude and sharpness of the overshoot occur at
lowest temperatures (0.05--0.2~K), while with increasing $T$ above
0.2~K the overshoot is smeared gradually. Overshoot is observed at almost all
$\nu$, with the exception of $\nu =2$ and 4, where overshoot is not observed
at all temperatures; for $\nu =5$ and 7, it is feebly marked only at
intermediate temperatures.

\begin{figure}
\includegraphics[width=7 cm]{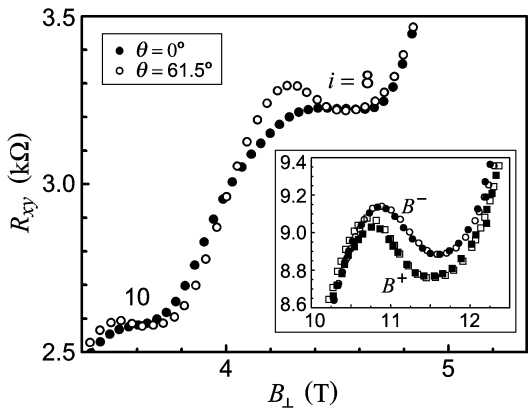}
\caption{$R_{xy}$ for $\nu = 8$ and 10 measured at $T = 0.33$~K as
a function of perpendicular component $B_\bot$ in tilted magnetic field
$(\phi = 61.5^\circ)$ in comparison with zero angle. The insert shows
the overshoot for $\nu = 3$ measured at $T = 0.33$~K when the
magnetic field was slowly tuned up and down (solid and open circles) in two
opposite directions ($B^+$ and $B^-$).}
\label{f.2}
\end{figure}

In Ref.~\cite{b06}, it was reported that in a Si/Si$_{1-x}$Ge$_x$ in tilted
magnetic fields, the overshoot at odd filling factor $\nu = 3$ increases
significantly at a tilting angle of around 69$^\circ$. This was interpreted
as being a consequence of the exchange enhancement of spin- and
valley-splitting attributed to band crossing of the cyclotron and
valley-split LB (proportional to the perpendicular component $B_\bot$) and
spin-split LB (proportional to the total field $B$). Our measurements show
that an increase of overshoot in tilted magnetic fields can also be observed
for even filling factors $\nu = 8$ and 10 (Fig.~\ref{f.2}), where the origin
of the adjacent LBs splitting differs in principle from the case $\nu = 3$.

Insert in Figure~\ref{f.2} shows the overshoot near $\nu = 3$ at
$T =0.33$~K measured for opposite directions of the perpendicular magnetic
field (marked as $B^+$ and $B^-$) and in the case when $B$ is tuned up and
down (solid and open circles). The coexistence of the curves as the magnetic
field is tuned up and down provides an evidence for the steady-state
phenomenon. Changing the direction of $B$ shows that the overshoot is
non-erasable after averaging and, therefore, cannot be explained by the
admixture of $\rho_{xx}$ caused by geometrical asymmetry in the sample.

In Ref.~\cite{b07}, it has been shown that in
two-dimensional SiGe hole gas, overshoots in $\rho _{xy}$ may occur as a
result of oscillations described by semi-classical theory and can be related
to the oscillations of longitudinal conductivity $\Delta \sigma _{xx}$
around the classical Drude conductivity: $\sigma _{xx}=n_{0}e\mu
_{0}/(1+\mu_{0}^{2}B^{2})$. Accordingly, the values of $\Delta \sigma _{xy}$
are obtained from oscillations around the classical curve $\sigma
_{xy}=\mu_{0}B\sigma _{xx}$. In accordance with this model, the amplitude of
the resistivity oscillations should be given by $\Delta \rho _{xy}=-\Delta
\rho_{xx}/2\mu _{0}B$. Our experimental data in $n$-type SiGe are in
disagreement with this relation: calculated $\Delta \rho _{xy}$ are too
small in comparison with the measured values which was mentioned also
in~\cite{b09}. This means that the above model cannot explain overshoot in
$n$-type SiGe.

\begin{figure}
\includegraphics[width=7 cm]{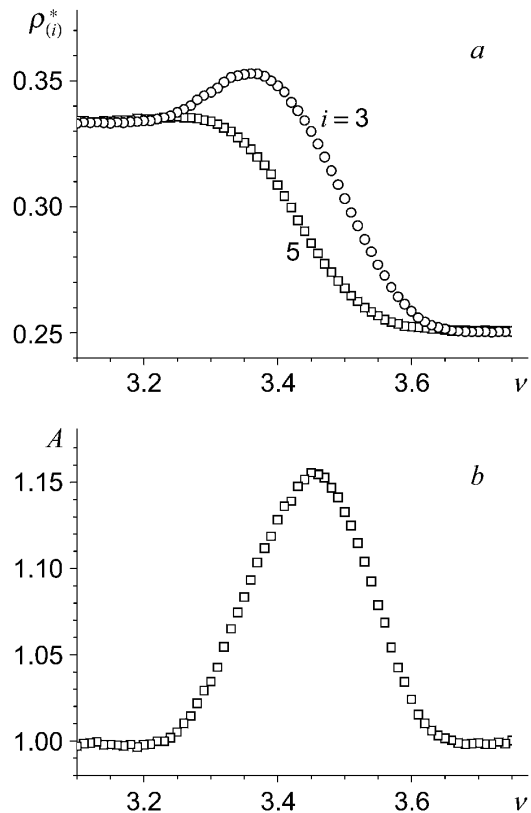}
\caption{(a) --- The normalized Hall resistance
$\rho_{(3)}^{\ast }=\rho^{\ast (4-3)}$ and rescaled
$\rho^{\ast (5-6)}=\rho _{(5)}^{\ast }$ as examples of experimental curves with
and without overshoot. (b) --- experimental ratio
$A(\nu)=\rho_{(3)}^{\ast }/\rho _{(5)}^{\ast}$ .}
\label{f.3}
\end{figure}

Let us compare the experimenal curves with and without overshoot.
Fig.~\ref{f.3}a shows the dependence of the normalized Hall resistivity
$\rho^{\ast(4-3)}\equiv R_{xy}(\nu )/(h/e^{2})$ for filling factor $\nu $
between $i=4$ and 3 where overshoot is clearly visible. This curve is labeled
as $\rho_{(3)}^{\ast }$. As an experimental curve without overshoot, we use the
dependence of $\rho ^{\ast (6-5)}$ between $i=6$ and 5. For comparison of
these two curves, we replot $\rho ^{\ast (6-5)}$ in the interval $\nu $
between $i=4$ and 3 using the following scaling procedure:
$\rho_{(4-3)}^{(6-5)}=[(\rho ^{\ast (6-5)}-1/6)(1/3-1/4)/(1/5-1/6)]+1/4$. This
curve is labeled as $\rho _{(5)}^{\ast }$. The experimental points of the
ratio $A(\nu )=\rho _{(3)}^{\ast }/\rho _{(5)}^{\ast }$ are plotted in
Fig.~\ref{f.3}b. One can see that the overshoot effect exists in the almost whole
interval between plateaus, not only in the region where $\rho _{xy}$ exceeds
$h/ie^{2}$, moreover, the maximal value of $A$ is achieved approximately in
the middle point between $\nu $ and $\nu +1$.

Observation of the overshoot for different materials and for different $\nu$,
where splitting of the adjacent LBs has different character, hints at the
common origin of this effect.

\begin{figure}[t]
\includegraphics[width=7 cm]{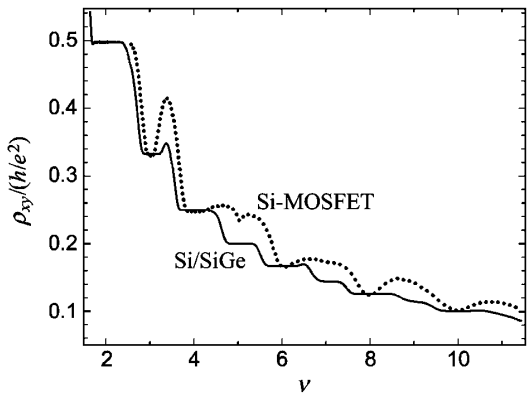}
\caption{Transversal Hall resistance $\rho _{xy}$ in units of
$h/e^2$ as a function of filling factor $\nu$ measured in Si/SiGe
($n=8.95\cdot 10^{15}$~m$^{-2}$, $\mu_0=8$~m$^2$/V$\cdot$s) and in
Si-MOSFET ($n=8.6\cdot 10^{15}$~m$^{-2}$, $\mu_0=1$~m$^2$/V$\cdot$s).
$T=50$~mK.}
\label{f.4}
\end{figure}

\begin{figure}
\vspace*{10mm}
\includegraphics[width=7 cm]{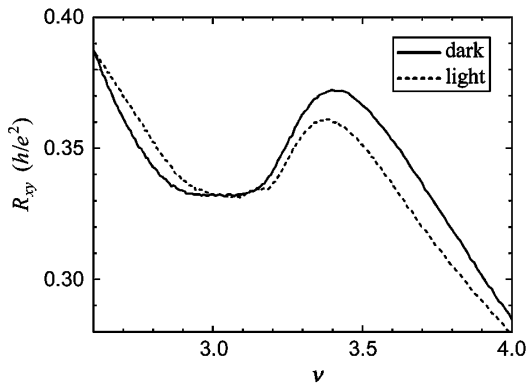}
\caption{$R_{xy}(\nu)$ in the vicinity of $\nu =3$ for $p$-type Si/SiGe
sample in dark (solid line) and after illumination (dashed line). $T=300$~mK.}
\label{f.5}
\end{figure}

We assume that the overshoot could be caused by existence of the second sort
of electrons with lower mobility which is determined at zero and weak magnetic
fields. In 3D classical conductivity, parallel contribution of two sorts of
carriers with different mobility leads to the overshoot in the measured
value of the Hall voltage even at strong magnetic fields \cite{b12} although
for 2D quantized conductivity, the classical approach is
inapplicable. However, we are not versed in theoretical calculation of
transversal resistance $\rho _{xy}$ in the QHE regime in the case
of two sorts of carriers. Carriers with lower mobility can be
originated, for example, from the barely localized states in the flanks of
the LBs that occupy the energy interval between
the delocalized states in the central part of LB and strongly localized
states in the tails of the band.

It turnes out that the above assumption is useful for undestanding of some
peculiarities of the overshoot phenomenon. For example, smearing of the
overshoot with increase of temperature (Fig.~\ref{f.1}) can be explained by
temperature-induced expansion of the delocalized states and corresponding
narrowing of the energy interval for barely localized states. Broadening of
LBs leads to the expansion of the barely localized states and therefore is
favorable for the overshoot observation. The broadening of LBs may be
caused, for example, by increase of overlapping of the adjacent LBs in
tilted magnetic fields (Fig.~\ref{f.2}) or by increase of disorder. This may
explain the reversible defect-induced appearance of the overshoot after
proton irradiation observed in a GaAs/AlGaAs sample~\cite{b13}. In Fig.~\ref{f.4}
the transversal Hall resistivity $\rho _{xy}$
in the investigated n-Si/SiGe heterostructure is
compared with a Si-MOSFET sample with close electron
concentration $n_{0}=8.6\cdot 10^{15}$~m$^{-2}$, but lower mobility
$\mu _{0}=1$~m$^2$/V$\cdot$s. One can see that in
the case of disordered 2D Si-MOSFET, the overshoot is strongly pronounced.

The next example of the influence of disorder is shown in Fig.~\ref{f.5}. In this
experiment, $p$-type Si/SiGe sample was investigated. Concentration of holes
and mobility measured at 1.5~K were $p_{0}=1.6\cdot 10^{15}$~m$^{-2}$
and $\mu _{0}=0.7$~m$^{2}$/V$\cdot$s. The transversal resistance $R_{xy}$
of this sample was measured in the dark and after illumination by a red LED
(LED current was 100~$\mu$A, duration 5~s). As a result of illumination, an
additional concentration of carriers was ''frozen'' and residual hole
concentration measured up to $1.8\cdot 10^{15}$~m$^{-2}$. It is known that
increase of concentration of ``frozen'' carriers leads to the more effective
screening of the random potential relief and therefore decreases disorder.
Fig.~\ref{f.5} shows $R_{xy}(\nu)$ in the vicinity of $\nu =3$  in dark (solid
line) and after illumination (dashed line) measured at $T=300$~mK. It is
seen that decrease of disorder results in decrease of the overshoot.

\acknowledgments We are thankful to S. Kravchenko for supplying us with
Si-MOSFET and p-Si/SiGe samples. I. S. and V. G. thank the Erick and Sheila Samson Chair of
Semiconductor Technology for financial support.

\end{document}